\documentclass{article}

\PassOptionsToPackage{numbers, compress}{natbib}
    \usepackage[preprint]{neurips_2020}

\usepackage[utf8]{inputenc} 
\usepackage[T1]{fontenc}    
\usepackage{hyperref}       
\usepackage{url}            
\usepackage{booktabs}       
\usepackage{amsfonts}       
\usepackage{nicefrac}       
\usepackage{microtype}      

\usepackage{graphics}       
\usepackage{arydshln}       
\usepackage{subcaption} 
\usepackage{graphicx}
\usepackage{amsmath}
\usepackage{amssymb}
\usepackage{xcolor}
\usepackage{enumitem}
\usepackage{wrapfig}
\usepackage{multirow}

\title{Perturbing Across the Feature Hierarchy to Improve Standard and Strict Blackbox Attack Transferability}

\author{%
  Nathan Inkawhich, Kevin J Liang,  Binghui Wang, Matthew Inkawhich, \\  \textbf{Lawrence Carin and Yiran Chen}\\
  Duke University\\
  \texttt{nathan.inkawhich@duke.edu} \\
}

\newcommand{\FDAfd}{\textit{FDA+fd}}
\newcommand{\fdanlxent}[1]{\textit{FDA}$^{(#1)}$\textit{+xent}}
\newcommand{\fdanl}[1]{\textit{FDA}$^{(#1)}$}

\begin{document}

\maketitle

\begin{abstract}
    We consider the blackbox transfer-based targeted adversarial attack threat model in the realm of deep neural network (DNN) image classifiers. Rather than focusing on crossing decision boundaries at the output layer of the source model, our method perturbs representations throughout the extracted feature hierarchy to resemble other classes. We design a flexible attack framework that allows for multi-layer perturbations and demonstrates state-of-the-art targeted transfer performance between ImageNet DNNs. 
    We also show the superiority of our feature space methods under a relaxation of the common assumption that the source and target models are trained on the same dataset and label space, in some instances achieving a $10\times$ increase in targeted success rate relative to other blackbox transfer methods. 
    Finally, we analyze why the proposed methods outperform existing attack strategies and show an extension of the method in the case when limited queries to the blackbox model are allowed.
\end{abstract}



\section{Introduction}
\vspace{-2mm}
The adversarial machine learning community has devised many ways to cause Deep Neural Networks (DNNs) to behave unexpectedly \cite{Szegedy2013IntriguingPO, Goodfellow2015ExplainingAH, Carlini2016TowardsET, MoosaviDezfooli2016DeepFoolAS, Madry2018TowardsDL}. However, the knowledge assumptions and threat models considered by an adversary are critical to attack success. In settings where access to and familiarity with the the target model is restricted, there is significant room for improving adversarial methods in terms of potency and efficiency, which is the central motivation for this work.

We focus on the blackbox transfer-based adversarial threat model for DNN image classifiers. In the standard case, blackbox means the attacker does not have access to the gradients of the target model and makes no assumptions about its architecture. Transfer-based indicates that adversarial examples are created by computing adversarial perturbations using a substitute whitebox model and then attacking the target blackbox model with the resulting examples, leveraging the notion of \textit{transferability} \citep{Papernot2016TransferabilityIM, Papernot2016PracticalBA, Tramr2017TheSO}. Within this threat model, our specific goal is targeted adversarial attacks, meaning the objective is to induce the target model to output a specific class. 
\begin{wrapfigure}{r}{.56\linewidth}
    \vspace{-4.5mm}
    \centering
        \includegraphics[width=.99\linewidth,trim={0 0 0 0},clip]{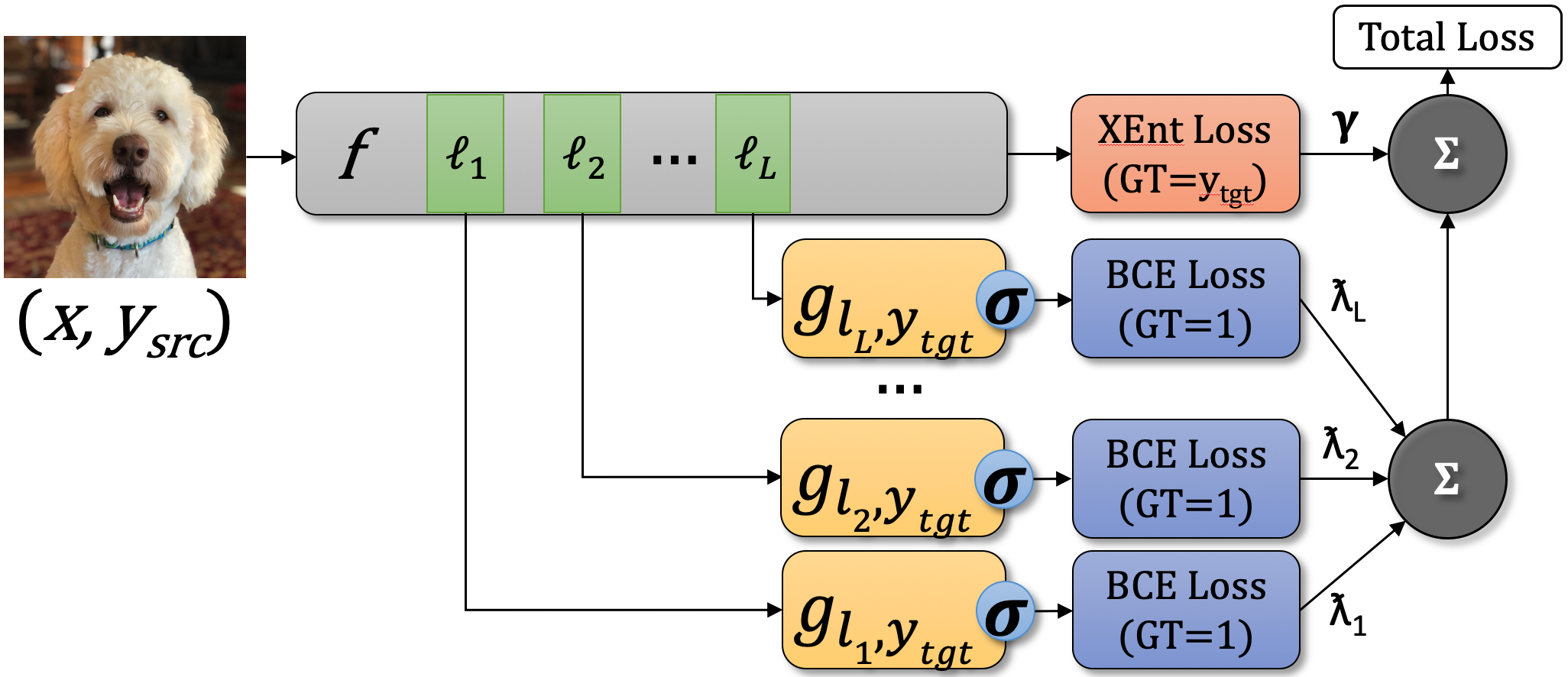}
    \vspace{-5mm}
    \caption{Visualization of forward pass construction to optimize our feature space attack objective.}
    \label{fig:forward-pass}
    \vspace{-10mm}
\end{wrapfigure}
These are significantly more challenging than un-targeted attacks, which seek to simply cause an incorrect prediction. In addition, we consider a set of more ``strict'' blackbox threat models in which we make varying degrees of assumptions about access to the blackbox model's training data distribution. This includes cases when the label spaces of the whitebox and blackbox models differ, and when there is \textit{zero} training data overlap.

A recent innovation in transfer attacks within the standard blackbox threat model is to craft perturbations based on intermediate layer representations, rather than simply optimizing the classification loss at the output layer \cite{RozsaGB17, Inkawhich_2019_CVPR, Inkawhich_2020_ICLR, lu2020enhancing}. An example of such a method is the Feature Distribution Attack (FDA)~\citep{Inkawhich_2020_ICLR}, which computes adversarial examples using intermediate feature distributions at a single layer of the whitebox model. In this work, we propose to significantly improve the FDA method by extending it into a more flexible framework to allow for perturbations across the intermediate feature space, including the output layer. Our method relies on modeling the layer-wise and class-wise feature distributions of the whitebox model via auxiliary networks. We then optimize adversarial noise using the auxiliary models from across the deep feature space. The critical observation in this work is that by enforcing that a perturbed sample ``looks-like'' a target class sample \textit{in multiple layers across the depth of the whitebox model}, the resulting sample is likely to have a much higher transfer rate than if we only considered a single intermediate layer or the output layer in isolation. For an intuition for how the attack works, consider Figure \ref{fig:forward-pass}, which shows the forward pass required to generate adversarial noise given a pre-trained DNN $f$. Each $g_{\ell,y_{tgt}}$ is a feature distribution model that estimates the probability that the layer $\ell$ feature map $f_\ell(x)$ is from a sample of class $y_{tgt}$, i.e. $p(y_{tgt}|f_\ell(x))$. For a chosen $y_{tgt}$ and set of layers, we accumulate the losses w.r.t. $y_{tgt}$ at each intermediate layer and the output layer. By optimizing the sum, we are noising $x$ with $\delta$ such that $x+\delta$ lies in high probability regions of the target class at several layers across feature space. We find that this method significantly improves transferability of the generated adversarial samples. 

To evaluate our methods, we consider both standard and strict blackbox transfer cases. In the standard case, our attacks show state-of-the-art targeted transfer performance between popular ImageNet-1K~\cite{Deng2009ImageNetAL} DNN models. In some cases, we improve the targeted success rate to over $55\%$, an absolute increase of about $50\%$ over traditional output-layer-based methods. In the strict blackbox scenario, we evaluate our method under three separate relaxations of the common transfer assumption that the source and target models are trained on the same data distribution and share a label space. The results show that our feature-based attacks are significantly more potent than attacks generated at the output layer in all three situations. In an effort to explain why our methods yield high transferability, we also analyze the effects of our attacks and show that they cause significantly more disruption in the intermediate space than competing methods. Finally, we show that the noise generated with our methods provides a more useful prior direction for query-based attacking methods that incorporate prior information. With relatively few queries to the blackbox model, our method can be enhanced to yield over $90\%$ targeted success rates.

\vspace{-3mm}
\section{Related work}
\vspace{-3mm}
In transfer-based blackbox adversarial attack research, many of the most popular methods build directly upon whitebox attacks \cite{Goodfellow2015ExplainingAH, Kurakin2016AdversarialML, Madry2018TowardsDL}, adding optimization tricks, regularization, and ensembling of whitebox models to boost transferability \cite{Kurakin2016AdversarialML, Papernot2016TransferabilityIM, Dong2018BoostingAA, Zhou2018TransferableAP, LiuCLS17, XieZZBWRY19}. The primary optimization goal of such methods is to cause the adversarial sample to cross the whitebox model's decision boundary through consideration of the classification loss. 
For example, \citet{Dong2018BoostingAA} includes a momentum term in the optimization step of \cite{Kurakin2016AdversarialML}; \citet{sgm_attack} weights the gradients through the residual connections and blocks differently; \citet{LiuCLS17} and \citet{Tramr2018EnsembleAT} use an ensemble of whitebox models; \citet{XieZZBWRY19} use diverse inputs for more robust noise; and \citet{SharmaDB19} and \citet{DongPSZ19} use low frequency constraints and translation-invariance respectively to generate un-targeted attacks for defended DNNs. These methods show reasonable un-targeted performance, but very limited targeted performance at scale, if any (\cite{DongPSZ19, Tramr2018EnsembleAT, sgm_attack, XieZZBWRY19} only consider un-targeted attacks).

There are also some recent methods that focus on improving transfer attacks by considering the feature space of the source model, to develop noise that is less overfit to the generating architecture. \citet{Zhou2018TransferableAP} develop a regularizer for a traditional un-targeted attack that encourages adversarial examples to have significantly different intermediate feature representations. \citet{HuangKGHBL19} adjusts an existing un-targeted adversarial example to have a larger effect in a pre-specified layer of feature space. \citet{RozsaGB17} and \citet{Inkawhich_2019_CVPR} develop targeted transfer attacks in feature space, representing the ``target'' as a single point in a single layer of the feature space. The objective functions then minimize the $L_2$ distance between a source and target point in the feature space. With this limited modeling of the target class, scalability to larger models and datasets proves difficult, and success is sensitive to the choice of the target sample. To improve targeted transfer performance at scale, \citet{Inkawhich_2020_ICLR} explicitly model the class-wise feature distributions at multiple layers of the whitebox model, so as to have a more descriptive representation of the target class when attacking. However, the method only considers attacking from a single layer (with no options for extensions), and the primary result is that some individual layers are better than others to transfer from. As is common in all of the aforementioned methods, results are only shown for transfer scenarios where the whitebox and blackbox models are trained on the same dataset. For comparison, we develop a flexible framework that allows for perturbations of multiple layers simultaneously, evaluate in novel cross-distribution settings, and show an integration of the method with a query-based attack.

\vspace{-2mm}
\section{Methodology}
\vspace{-2mm}
\textbf{Notation.}\ We follow the notation conventions of \citet{Inkawhich_2020_ICLR} for feature distribution modeling. To capture the layer-wise and class-wise feature distributions of a given pre-trained DNN $f$, we first specify a set of layers $\mathcal{L}=\{\ell_1,...,\ell_J\}$ and classes $\mathcal{C}=\{c_1,...,c_K\}$ of interest. Using the training dataset of $f$, we train binary classification auxiliary models on the feature maps at the layers in $\mathcal{L}$ for the classes in $\mathcal{C}$. Importantly, all weights of $f$ stay fixed throughout this process. Let $f_\ell(x)$ be the layer $\ell$ feature map of $f$ for input image $x$. Then, $g_{\ell,y}$ is a binary auxiliary model that inputs $f_\ell(x)$, and outputs the probability that the feature map is from an input of class $y$ (i.e., $g_{\ell,y}(f_\ell(x)) = p(y | f_\ell(x))$). 
Note, the training of the necessary auxiliary models is done prior to the attacking process, and requires only a pre-trained model $f$ and its corresponding training dataset. It is also worth noting that each auxiliary model is small and individually very inexpensive to train (see Appendix B).
Finally, let $f(x)$ be the predicted probability distribution over the set of classes on which $f$ was trained, and $F(x) = \mathrm{argmax}\ f(x)$ be the classification prediction.


\textbf{Preliminaries.}\ Once trained, we can use the modeled feature distributions to attack. Given our interest in targeted attacks, we start from a source image $x$ initially classified as $y_{src}$ (i.e. $F(x)=y_{src}$) and generate a perturbation $\delta$ such that $F(x+\delta)=y_{tgt}$ for a pre-specified target class $y_{tgt}$, with $y_{src} \neq y_{tgt}$. To keep the noise approximately imperceptible and thus adversarial, for the following objectives we constrain the $L_p$ norm of $\delta$, such that $||\delta||_p\leq \epsilon$, and enforce that the perturbed input's pixel values exist in $[0,1]$.

\textbf{\textit{FDA}.}\ Our proposed attack framework builds on \FDAfd, the top performing variant of \cite{Inkawhich_2020_ICLR}, which is noted here as \textit{FDA} and described as
\begin{equation}
    \max_{{\delta}}  p(y=y_{tgt}|f_\ell(x+\delta)) + \eta \frac{\left\Vert f_\ell(x+\delta) - f_\ell(x) \right\Vert _2}{\left\Vert f_\ell(x) \right\Vert _2}. 
    \label{eqn:fda}
\end{equation}
This method has two main components, both of which are optimized at a \textit{single} pre-specified layer $\ell$. The first component is the feature distribution piece, which takes the form of $p(y=y_{tgt}|f_\ell(x+\delta))$. By maximizing this probability through optimizing $\delta$, we are perturbing the input image such that the layer $\ell$ feature map $f_\ell(x+\delta)$ lies in a high probability region of the target class in feature space. In other words, the perturbed feature map resembles a feature map that will be classified as the target class. The second component is the feature disruption term, which enforces that the feature map of the perturbed input is significantly different than the feature map of the original input.

\textbf{\textit{FDA+xent}.}\ A key contribution in this work is the extension of (\ref{eqn:fda}) to include \textit{multi-layer} information. One way to do so is to incorporate the whitebox model's output prediction as part of the attack objective. Let $H(f(x),y)$ be the standard cross-entropy loss between the predicted probability distribution $f(x)$ and the target distribution $y$ (commonly a one-hot distribution). We include this term in the \textit{FDA} objective to create \textit{FDA+xent}, as described by
\begin{equation}
    \max_{{\delta}} p(y=y_{tgt}|f_\ell(x+\delta)) + \eta \frac{\left\Vert f_\ell(x+\delta) - f_\ell(x) \right\Vert _2}{\left\Vert f_\ell(x) \right\Vert _2} - \gamma H(f(x+\delta),y_{tgt}),
\end{equation}
where $\gamma>0$, weighting the contribution of cross-entropy term. This attack objective optimizes the noise such that the layer $\ell$ feature map is in a high-probability region of the target class, and that the output prediction of the whitebox model is of the target class. 
Addition of the cross-entropy term is motivated by the correlation results from \citet{Inkawhich_2020_ICLR}, which show that the probability distribution over the classes, as measured in the ``optimal transfer layer,'' has low correlation with the probability distribution over the classes at the output layer. 
This indicates that the vanilla \textit{FDA} in (\ref{eqn:fda}) does not reliably create targeted examples for the whitebox model, which may limit the effectiveness of the attack; we explore this further in Section \ref{sec:disruption}.

\textbf{\textit{FDA}$^{(N)}$.}\ To further generalize the multi-layer framework, we extend the objective function to allow for optimization in \textit{multiple intermediate layers}. We call this new attack \fdanl{N}, with objective: 
\begin{equation}
    \max_{{\delta}} \sum_{\ell \in \mathcal{L}} \lambda_\ell \bigg [ p(y=y_{tgt}|f_\ell(x+\delta)) + \eta \frac{\left\Vert f_\ell(x+\delta) - f_\ell(x) \right\Vert _2}{\left\Vert f_\ell(x) \right\Vert _2} \bigg ], 
\end{equation}
where $\sum_{\ell \in \mathcal{L}} \lambda_\ell = 1$, $\lambda_\ell > 0$, and the \textit{N} in the name refers to the number of layers in $\mathcal{L}$. Note that for $N=1$, the attack objective is the same as (\ref{eqn:fda}). Both the \textit{+xent} and \textit{multi-intermediate-layer} extensions significantly increase the capability and flexibility of the method, while maintaining the intuition of attacking primarily based on feature space information. These two extensions can also be straightforwardly composed to form a \fdanlxent{N} attack. 

\textbf{Optimization.}\ The final step in the methodology is the optimization of the above objective functions. Practically, we leverage the autograd feature of PyTorch by assembling the forward pass of the attack as shown in Figure \ref{fig:forward-pass}. This step involves connecting the required auxiliary models to the corresponding layers of $f$ and computing the final loss with proper weighting. We then use an iterative projected gradient descent with momentum procedure \citep{Dong2018BoostingAA} to apply the adversarial noise to the input while maintaining the norm and natural image constraints. 
\vspace{-2mm}
\section{Experiments}
\vspace{-2mm}
We evaluate the multi-layer \textit{FDA} framework in a variety of settings. In Section \ref{sec:top-inet-trans}, we consider the standard blackbox transfer case, where knowledge of the target model's training dataset is assumed, and test transferability between ImageNet-1K models. We examine single-source single-target model transfers, ensemble transfers, distal adversaries, and analyze why and how the attacks work through the lens of feature disruption. In Section \ref{sec:cross-dist} we evaluate transferability in cross-distributional settings, where the source and target model are trained on different data distributions and label spaces. Finally, in Section \ref{sec:query-extension} we show an extension of the method when limited queries to the target model are allowed during attack generation.

\vspace{-2mm}
\subsection{ImageNet transfer experiments} \label{sec:top-inet-trans}
\vspace{-2mm}
\textbf{Experimental Setup.}\ In our analysis, we use the following ImageNet-1K models from the PyTorch Model Zoo: ResNet-34/50/101/152 (RN34, RN50, RN101, RN152) \citep{He2016DeepRL}; VGG16bn and VGG19bn \citep{Simonyan2015VeryDC}; MobileNetv2 (MNv2) \citep{SandlerHZZC18}; and DenseNet-121/161/201 (DN121, DN161, DN201) \citep{Huang2017DenselyCC}. The notation RN50$\rightarrow$DN121 means attacks are generated on a RN50 whitebox (source) model and transferred to a DN121 blackbox (target) model. The primary metrics of attack success are the error rate and targeted success rate (tSuc) in the blackbox model, where all clean samples are correctly classified by both the source and target models. For the attack settings, we use the common $L_\infty \ \epsilon=16/255$ and $\mathrm{perturb\_iters}=10$. The architecture of each auxiliary model $g_{\ell,y}$ is a simple Conv-Conv-FC scheme, which is significantly more parameter efficient than the FC-FC in \cite{Inkawhich_2020_ICLR}. 

To find the best combinations of layers for \fdanl{N} attacks, we use an iterative greedy search on a held out part of the ImageNet validation set and find value in including up to 5 layers. Crucially, feature space attacks have been shown to be \textit{blackbox model agnostic}: the optimal transfer layer for a whitebox model does not change for different target blackbox models \citep{Inkawhich_2019_CVPR, Inkawhich_2020_ICLR}. For this reason, we find the optimal layer sets for the RN50 and DN121 whitebox models only once, in the RN50$\rightarrow$DN121 and DN121$\rightarrow$RN50 transfer scenarios, respectively. When attacking any other blackbox model, we use the previously found layer combinations. Practically, an attacker may have two models in a sandbox environment, where one is treated as a whitebox and the other a blackbox. They may then find the optimal layer settings in this sandbox environment, which can be used to attack any other blackbox of interest. The layer decoding scheme and the sequence in which the layers are added to the attack is shown in Appendix A. Finally, we weight all intermediate layers equally, i.e. $\lambda_\ell = 1/N$. Additional details regarding experimental setup are in Appendix B.

\vspace{-2mm}
\subsubsection{Transfer results in standard blackbox settings} \label{sec:Inet_transfer}
\vspace{-2mm}
We perform single-source single-target model transfers for RN50$\rightarrow$\{DN121, VGG16bn, RN152, MNv2\} and DN121$\rightarrow$\{RN50, VGG16bn, DN201, MNv2\}. For baselines, we compare against the popular TMIM \citep{Dong2018BoostingAA} and TMIM+SGM \citep{sgm_attack}, both of which search for adversarial directions using \textit{only output layer information} of the whitebox model. \fdanl{1} \citep{Inkawhich_2020_ICLR} is also considered a baseline. In an effort to emphasize the transferability of the noise generation technique when the blackbox model architecture may be unknown, we favor settings where the source and target model are from different architectural families to avoid potential biases.  

The procedure for measuring the transfer results in Table \ref{tab:full-inet-transfer-table} is as follows. For each source-target model pair, we randomly select 15,000 source images from the ImageNet validation set for which both models initially predict the correct source label. For each source sample, we then randomly select 5 target labels from the ImageNet label set, and execute a targeted attack towards each. Thus, every table entry is an average computed over 75,000 targeted attacks.
Both error rate and targeted success rate, as measured in the blackbox model, are reported as "error / tSuc".

\begin{table}[t]
\caption{Full-ImageNet Transfer Results (notation = error / tSuc, $\epsilon=16/255$)}
\label{tab:full-inet-transfer-table}
\resizebox{\columnwidth}{!}{
\begin{tabular}{c||cccc|cccc}
\toprule
                                                   & \multicolumn{4}{c|}{Whitebox Model = RN50}            & \multicolumn{4}{c}{Whitebox Model = DN121}            \\ \hline
Attack \textbackslash{}\textbackslash\ Target      & DN121       & VGG16bn     & RN152       & MNv2        & RN50        & VGG16bn     & DN201       & MNv2        \\ \hline \\[-2.5mm]
TMIM                                               & 44.7 / 3.0           & 48.7 / 1.5           & 41.4 / 3.2           & 54.9 / 0.1           & 46.2 / 2.3  & 49.6 / 1.3  & 44.6 / 5.5  & 58.5 / 1.0  \\
TMIM+SGM                                           & 47.4 / 4.5           & 51.9 / 2.4           & 43.1 / 4.2           & 61.1 / 2.2           & 54.8 / 4.7  & 53.8 / 2.8  & 51.4 / 10.0 & 66.1 / 3.2 \\ [0.5mm] \cdashline{1-9}  \\[-2mm]
FDA$^{(1)}$                                        & 90.3 / 19.4          & 86.6 / 13.6          & 90.8 / 17.1          & 84.7 / 7.1           & 91.8 / 20.4 & 90.6 / 20.6 & 94.8 / 35.3 & 88.4 / 10.7 \\
FDA$^{(2)}$                                         & 93.1 / 28.4          & 91.5 / 22.9          & 92.7 / 22.0          & 89.8 / 10.7          & \textbf{95.1} / 22.3 & \textbf{94.7} / 26.3 & 96.6 / 38.9 & \textbf{93.0} / 12.4 \\
FDA$^{(3)}$                                         & 94.1 / 31.0          & \textbf{93.0} / 26.1 & 93.2 / 22.9          & \textbf{91.0} / 11.8 & 95.1 / 22.2 & 94.7 / 26.0 & \textbf{96.7} / 37.1 & 92.8 / 12.6 \\
FDA$^{(4)}$                                         & \textbf{94.2} / 38.1 & 92.4 / 30.7          & \textbf{93.6} / 30.6 & 90.2 / 14.9          & 92.9 / 43.7 & 92.4 / 42.6 & 96.3 / 67.5 & 88.9 / 20.9 \\
FDA$^{(5)}$                                         & 94.2 / 38.6          & 92.7 / 31.0          & 93.5 / 29.7          & 90.9 / 15.1          & 93.8 / 44.4 & 93.4 / 44.4 & 96.6 / 68.2 & 90.6 / 22.4 \\[0.5mm] \cdashline{1-9}  \\[-2mm]
FDA$^{(1)}$+xent                                    & 84.4 / 40.3          & 80.9 / 24.9          & 85.0 / 42.3          & 78.6 / 13.2          & 88.8 / 37.4 & 87.8 / 32.7 & 93.3 / 65.3 & 85.1 / 16.9 \\
FDA$^{(2)}$+xent                                    & 89.0 / 51.7          & 86.8 / 37.2          & 88.2 / 48.1          & 84.3 / 18.7          & 92.7 / 40.5 & 92.5 / 40.5 & 95.5 / 70.3 & 90.1 / 19.4 \\
FDA$^{(3)}$+xent                                    & 90.2 / 55.2          & 88.3 / 41.6          & 89.2 / 49.2          & 85.6 / 21.1          & 92.9 / 41.7 & 92.5 / 41.8 & 95.7 / 70.6 & 90.0 / 20.4 \\
FDA$^{(4)}$+xent                                    & 90.5 / 57.3          & 88.2 / 42.8          & 89.7 / \textbf{53.2} & 85.4 / 22.4          & 91.3 / 49.4 & 90.9 / 46.2 & 95.3 / 76.5 & 87.1 / 22.6 \\
FDA$^{(5)}$+xent                                    & 90.9 / \textbf{57.9} & 88.8 / \textbf{43.5} & 89.7 / 51.6          & 86.4 / \textbf{22.9} & 92.2 / \textbf{50.1} & 92.1 / \textbf{48.0} & 95.6 / \textbf{77.1} & 88.8 / \textbf{24.4} \\ \bottomrule
\end{tabular}
}
\vspace{-5mm}
\end{table}

Our first major observation is that the \textit{+xent} component alone gives large gains in tSuc. If we compare \fdanl{1} to \fdanlxent{1}, in several cases it leads to a tSuc increase of over $20\%$; for example, in RN50$\rightarrow$DN121 tSuc increases from $19.4\%$ to $40.3\%$. Further, we observe \textit{+xent} consistently helps, regardless of how many layers are used. The largest performance gain observed by \textit{+xent} is in the DN121$\rightarrow$DN201 case, where in the $1$ attack layer case, tSuc increases by $30\%$. The next result of interest is the performance gain by adding multiple intermediate layers (\fdanl{N}, $N>1$). As $N$ is increased, tSuc nearly doubles in all cases. For perspective, when we compare these results to the baseline \textit{TMIM}-based methods, the targeted success rate in many cases improves by more than $10\times$.

We also observe that \textit{+xent} and \textit{multi-intermediate-layer} are complementary components: in all transfer scenarios the most powerful targeted attacks arise from using both. Specifically, consider the \fdanlxent{5} attack, which is the most effective in all but one case. When VGG16bn is the target model, on average this attack outperforms \textit{TMIM+SGM} by $37\%$ error / $43\%$ tSuc and \fdanl{1} by about  $2\%$ error / $29\%$ tSuc. An interesting observation is that the \textit{+xent} term harms error performance, albeit less relevant because our focus is targeted attacks, and the methods have been optimized for tSuc. However, in all cases the best \fdanl{N} method nearly doubles the error induced by \textit{TMIM+SGM}. 

\vspace{-2mm}
\subsubsection{Comparison to ensemble methods}
\vspace{-2mm}
\begin{wrapfigure}{r}{.45\linewidth}
\vspace{-8mm}
\centering
    \includegraphics[width=\linewidth,trim={10 0 30 0},clip]{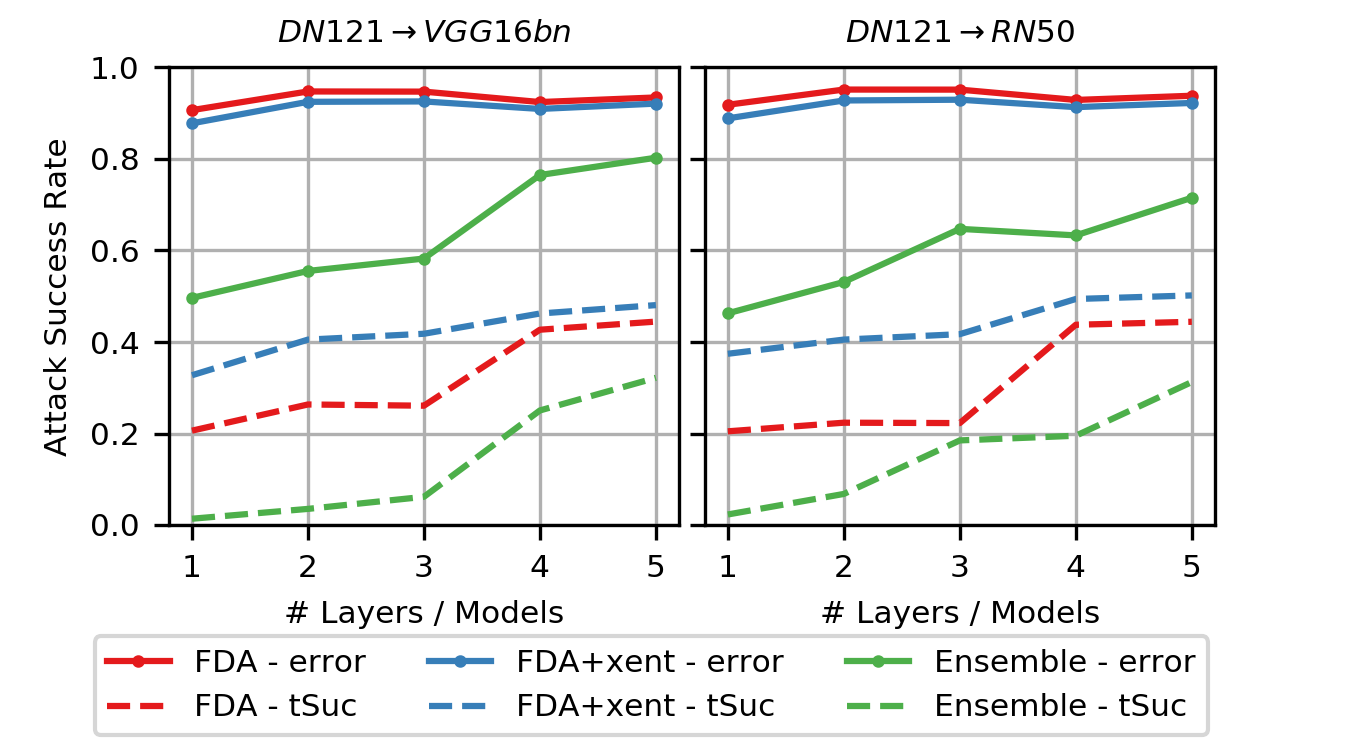}
    \vspace{-6mm}
    \caption{FDA versus ensemble attacks.}
    \label{fig:ensemble-comparison}
    \vspace{-4mm}
\end{wrapfigure}

Ensemble-based approaches \citep{LiuCLS17} form a separate family of methods to generate transferable adversarial examples, deviating from our previous single-source model assumption.
In this setting, the attacker trains an ensemble of models (with whitebox access to each) and generates noise based on the output of the ensemble with traditional attacking techniques, e.g., \textit{TMIM}. A natural iterate in ensemble methods is growing the numbers of models. We compare the effect of adding a layer within the \fdanlxent{N} framework versus adding a model to the generating ensemble using the same attack settings as in the previous experiment. To avoid conflicts with models already in use, in every transfer case we use the following sequence for adding models to the ensemble: [whitebox, DN161, RN101, VGG19bn, RN34]. For example, in the RN50$\rightarrow$DN121 transfer case, using a 3-model ensemble means generating noise from RN50, DN161, and RN101.

The results of these experiments are shown in Figure \ref{fig:ensemble-comparison}, with extensions in Appendix C. The \textit{FDA} methods hold a wide margin over ensemble methods in almost all cases for both error and tSuc, especially \fdanlxent{N}. Also, although adding models to the whitebox ensemble tends to increase attack performance (as expected), large jumps in the ensemble's performance usually occur when a model from the same family as the blackbox model is added. For example, in the DN121$\rightarrow$VGG16bn case, a performance spike happens when the fourth model gets added, which is the VGG19bn model. Similarly, in the DN121$\rightarrow$RN50 case, the ensemble method has the greatest increases when the third and fifth models are included, which are the RN101 and RN34 models, respectively.

\vspace{-2mm}
\subsubsection{Distal transfers}
\vspace{-2mm}
\begin{wrapfigure}{r}{.30\linewidth}
    \vspace{-8mm}
    \center\includegraphics[width=.99\linewidth,trim={0 0 0 0},clip]{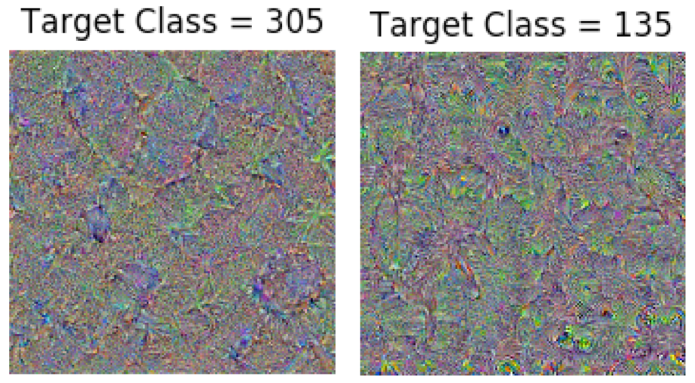}
    \vspace{-6mm}
    \caption{Samples of distals.}
    \label{fig:distal-samples}
    \vspace{-4mm}
\end{wrapfigure}

Distal adversarial examples are generated by starting from random noise and optimizing for high prediction probability for some target class~\cite{NguyenYC15}.
Transferring distals between ImageNet models in the standard blackbox setting can make for an interesting test of transferability.
Although distals are not our main focus, they provide a compelling future direction to study, as they depend less on the source image/class than standard transfers. This eliminates one axis of variability that may better isolate the quality of the attacking algorithm. For these tests, we do not constrain $\epsilon$ and increase the perturbation budget to allow each attack to perturb for $200$ iterations. 

\begin{wraptable}{r}{.45\linewidth}
    \vspace{-4mm}
    \centering
    \caption{Distal transfer tSuc rates (top1 / top5)}
    \vspace{-2.5mm}
    \label{tab:distal-table}
    \resizebox{\linewidth}{!}{
    \begin{tabular}{ccc}
    \toprule
    attack                   & RN50$\rightarrow$DN121 & RN50$\rightarrow$VGG16bn \\ \hline
    TPGD                    & 10.1 / 21.0             & 6.1 / 13.4              \\
    FDA$^{(1)}$+xent             & 46.9 / 68.1             & 30.6 / 50.6              \\
    FDA$^{(2)}$+xent             & 58.6 / 78.8             & 38.0 / 62.5              \\
    FDA$^{(3)}$+xent             & 64.0 / 83.6             & 44.6 / 69.9              \\
    FDA$^{(4)}$+xent             & \textbf{65.8} / \textbf{85.3}             & \textbf{48.1} / \textbf{73.0} \\ \bottomrule           
    \end{tabular}
    }
    \vspace{-4mm}
\end{wraptable}

The results are shown as top-1 / top-5 tSuc rates, as averaged over 4,000 distals, each optimized towards a random target class. We use \textit{TPGD} \citep{Madry2018TowardsDL} as our baseline instead of \textit{TMIM}, as momentum in the optimizer empirically harms performance in distal generation.
Table \ref{tab:distal-table} shows that results of the distal transfers align with the previous trends, and sample distals are shown in Figure \ref{fig:distal-samples}. Distals generated with the \fdanlxent{N} method are significantly more transferable. On average, the \fdanlxent{4} causes increases of $48\%$ top-1 / $62\%$ top-5 tSuc over the baseline method.

\vspace{-2mm}
\subsubsection{Analysis of \textit{multi-intermediate-layer} and \textit{+xent}} \label{sec:disruption}
\vspace{-2mm}
To analyze the effects of the \textit{multi-intermediate-layer} and \textit{+xent} components, we perform a feature disruption analysis \citep{Inkawhich_2020_ICLR}. Disruption measures the effect of an adversarial perturbation on the feature space of a model, w.r.t. the target class. In this setting, we measure the disruption caused by a targeted attack, at layer $\ell$ of model $f$ as:
$\mathrm{disruption_\ell} = p(y=y_{tgt}|f_\ell(x+\delta)) - p(y=y_{tgt}|f_\ell(x))$.
$f$ can be either a whitebox or blackbox model and the attack that generates $\delta$ may be any method. Figure \ref{fig:feature-disruption} shows the disruption caused by several attacks for RN50$\rightarrow$DN121 (see Appendix D for more). 

\begin{wrapfigure}{r}{.45\linewidth}
    \vspace{-5mm}
    \centering
        \includegraphics[width=.99\linewidth,trim={50 0 317 0},clip]{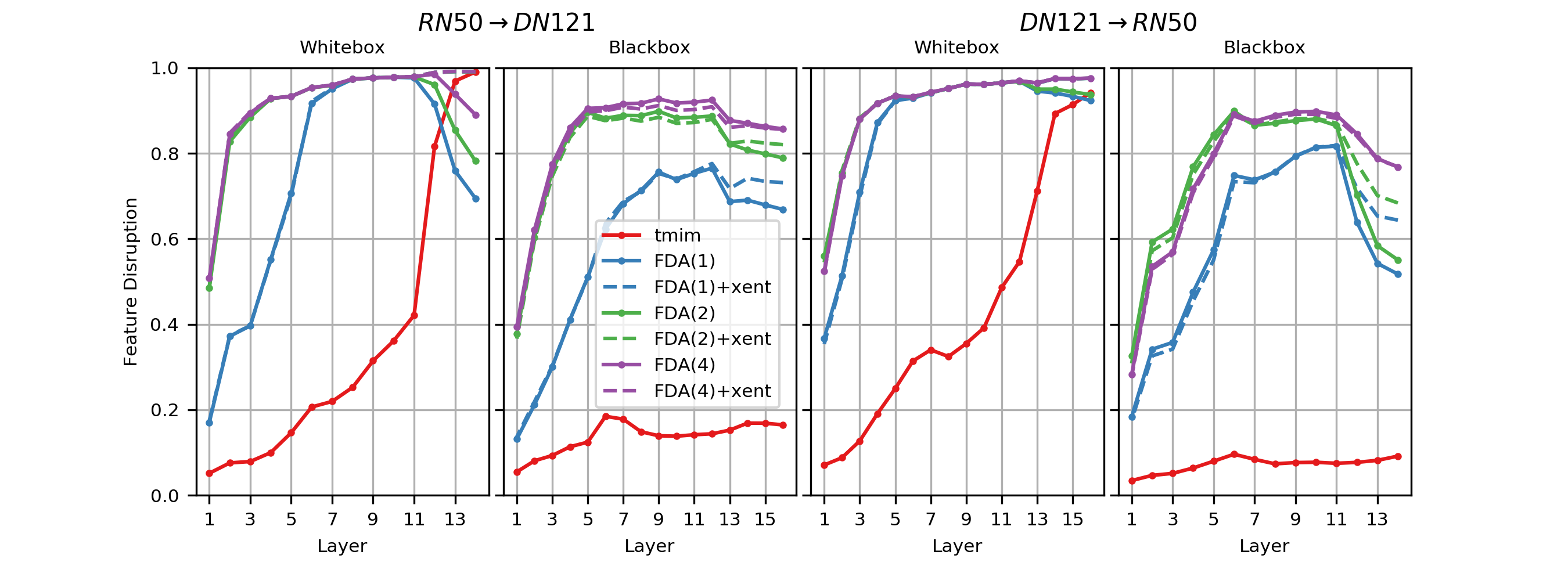}
    \vspace{-6mm}
    \caption{Disruption of features caused by transfer attacks.}
    \label{fig:feature-disruption}
    \vspace{-3mm}
\end{wrapfigure}

Firstly, the \textit{TMIM} attack is shown to have drastically different effects in the whitebox and blackbox models. Since the noise is generated specifically for the output layer of the whitebox, it only causes high disruption in the last few layers of the whitebox, and very little disruption in the blackbox. For the \fdanl{1} method, we already see a significantly different pattern, namely a much higher disruption in the intermediate layers of both models. Now, consider the effects of adding multiple intermediate layers. As we look from \fdanl{1} to \fdanl{2} to \fdanl{4}, the disruption is noticeably increased in the earlier and later layers, which is expected as the added intermediate layers are both earlier and later than the first layer. Finally, consider the \textit{+xent} component. Without it, the \fdanl{N} methods have sharp decreases in the last few layers, despite high intermediate disruption. The inclusion of this term significantly boosts disruption in the final layers of both the whitebox and blackbox. Given the large effects of our feature space methods throughout the models, we say that the methods attack along the feature hierarchy.

\vspace{-2mm}
\subsection{Cross-distribution experiments} \label{sec:cross-dist}
\vspace{-2mm}
An often overlooked (but critical) detail in standard blackbox transfer attacks is the underlying assumption that the whitebox and blackbox models are trained on the exact same dataset \citep{Dong2018BoostingAA, LiuCLS17, sgm_attack, Inkawhich_2019_CVPR, Inkawhich_2020_ICLR, Madry2018TowardsDL, Tramr2017TheSO}. This implies that the adversary can find such a pre-trained model in the open-source domain or actually has the blackbox model's dataset and can train their own whitebox model. In many cases, the dataset used to train the targeted model may not be publicly available due to proprietary~\cite{Graepel2010,Sun2017}, security~\cite{Liang2018, Sigman2020}, or privacy~\cite{USCensus2018, Ribli2018} issues. Making too strong of an assumption about the availability of the target model's data may lead to overly optimistic transfer results.
Here, we evaluate the performance of transfer attacks under several more ``strict'' blackbox threat models, where the whitebox model is not trained on the exact same data distribution or label space as the blackbox model. We hypothesize that it is more challenging to create transferable adversarial examples if the source and target models are not trained on identical data. 

We are primarily concerned with three ``cross-distribution'' transfer scenarios:
\vspace{-3mm}
\begin{enumerate}[leftmargin=*]
    \itemsep -0.75mm
    \item \textit{No training data overlap between the whitebox and blackbox, but significant label space overlap.}
    \item \textit{The blackbox is trained on a subset of the whitebox's training dataset and label space.}
    \item \textit{The whitebox is trained on a subset of the blackbox's training dataset and label space.}
    \vspace{-2.5mm}
\end{enumerate}

These three cases represent several common situations that may be encountered in reality. For example, scenarios 1 and 3 may arise when the attacker is generally aware of the types of classes that the target model is trained on, but has to go out and collect their own data for training the source model. Scenario 2 may arise when the attacker has access to a large model with a high variety of classes, and wishes to attack a smaller target model designed for a subset of the classes.

\textbf{Experimental setup.}\ To accommodate these scenarios, we create three new ``Restricted-ImageNet'' datasets (RINet), which are subsets of ImageNet-1k (Full-INet). 
We design \textit{RINet-A} and \textit{RINet-B} to each be 15-class datasets, sharing 10 label classes while each having 5 unique label classes.
Each of the classes in \textit{RINet-A} and \textit{RINet-B} consist of WordNet~\cite{miller1998wordnet} super-classes aggregating a related subset of ImageNet-1K classes. 
Critically, the classes chosen to comprise each of the shared classes of \textit{RINet-A} and \textit{RINet-B} are disjoint.
For example, the ``bird'' class of RINet-A is composed of the ImageNet-1k classes: \texttt{[10:`brambling', 13:`junco', 16:`bulbul', 17:`jay',  18:`magpie']}, while RINet-B contains \texttt{[12:`housefinch', 14:`indigobunting', 15:`robin', 19:`chickadee',  20:`waterouzel']}.
As a result, \textit{RINet-A} and \textit{RINet-B} have \textit{zero training data overlap}.
The \textit{RINet-C} dataset is comprised of 20 classes and is meant to be a bigger subset of Full-INet than RINet-A and RINet-B. We train RN50 and DN121 models on each of the RINet datasets, then train the auxiliary models necessary to attack with the \fdanlxent{N} framework. For more details regarding setup and the dataset splits, see Appendix E. Hereafter, the notation RN50-A indicates the RN50 model trained on RINet-A.

\begin{table}[b]
\vspace{-6mm}
\caption{Transfer results for cross-distribution tests (notation = error / tSuc)}
\label{tab:cross-distribution-results-table}
\resizebox{\columnwidth}{!}{
\begin{tabular}{c||ccc|ccc|ccc}
\toprule
                & \multicolumn{3}{c|}{Target Model = DN121-B}          & \multicolumn{3}{c|}{Target Model = DN121-A}          & \multicolumn{3}{c}{Target Model = DN121-Full} \\ \hline
Attack \textbackslash{}\textbackslash\ Source    & RN50-A       & RN50-B        & RN50-Full        & RN50-A        & RN50-B        & RN50-Full        & RN50-A          & RN50-B        & RN50-C        \\ \hline \\[-2.5mm]
TMIM            & 37.8 / 11.7 & 33.5 / 21.4 & 16.4 / 3.3   & 29.9 / 19.2 & 34.8 / 11.4 & 12.6 / 2.7  & 29.2 / 0.9   & 32.0 / 1.0  & 35.3 / 2.7  \\
TMIM+SGM        & 43.0 / 16.6 & 38.6 / 27.8 & 19.2 / 5.2   & 36.6 / 27.0 & 38.3 / 14.6 & 16.2 / 4.6  & 30.4 / 1.2   & 32.9 / 1.4  & 37.8 / 3.3    \\[0.5mm] \cdashline{1-10}  \\[-2mm]
FDA$^{(1)}$      & 67.7 / 45.8 & 70.9 / 60.8 & 38.6 / 21.0  & 66.0 / 57.3 & 67.0 / 40.8 & 34.8 / 18.9 & 53.8 / 6.9   & 58.1 / 8.1  & 75.8 / 14.1 \\
FDA$^{(2)}$      & 72.3 / 54.1 & 75.5 / 68.0 & 53.9 / 37.1  & 72.5 / 65.9 & 71.3 / 48.4 & 52.4 / 36.9 & 59.3 / 9.4   & 61.5 / 9.2  & 80.6 / 16.7 \\
FDA$^{(3)}$      & 73.9 / 58.0 & 77.4 / 72.4 & 58.3 / 42.1  & 76.9 / 72.8 & 70.7 / 50.4 & 57.0 / 42.3 & 55.2 / 11.1  & 57.4 / 10.7 & 77.9 / 22.1 \\
FDA$^{(4)}$      & 76.1 / 60.7 & 79.4 / 74.9 & 56.0 / 41.5  & 79.1 / 75.3 & 73.0 / 53.2 & 55.0 / 42.1 & 58.1 / 12.1  & 60.4 / 11.4 & 80.1 / 22.5 \\
FDA$^{(5)}$      & 76.5 / 61.9 & 79.7 / 75.4 & \textbf{59.2} / \textbf{45.5}  & 79.4 / 75.9 & 73.5 / 54.0 & \textbf{58.4} / \textbf{45.5} & 58.8 / 12.2  & 60.7 / 11.7 & 80.5 / 22.4 \\[1mm] \cdashline{1-10} \\[-2mm]
FDA$^{(1)}$+xent & 70.2 / 50.1 & 75.6 / 69.4 & 31.7 / 17.2  & 72.0 / 65.8 & 69.0 / 45.6 & 28.5 / 16.0 & 55.0 / 7.9   & 58.3 / 9.0  & 76.1 / 16.4  \\
FDA$^{(2)}$+xent & 74.5 / 58.3 & 79.8 / 75.2 & 43.9 / 29.8  & 77.2 / 73.0 & 73.3 / 53.0 & 42.4 / 30.7 & \textbf{60.4} / 10.5  & \textbf{62.1} / 10.7 & \textbf{81.2} / 19.4 \\
FDA$^{(3)}$+xent & 74.3 / 59.3 & 79.6 / 76.1 & 47.0 / 33.3  & 78.7 / 75.4 & 71.7 / 52.4 & 46.1 / 35.0 & 55.6 / 11.5  & 57.9 / 11.4 & 78.3 / 23.8  \\
FDA$^{(4)}$+xent & 76.5 / 62.2 & 81.3 / 78.0 & 45.9 / 32.7  & 80.8 / 77.7 & 74.1 / 55.3 & 44.7 / 34.0 & 58.7 / 12.6  & 61.0 / 12.1 & 80.6 / \textbf{24.3} \\
FDA$^{(5)}$+xent & \textbf{77.0} / \textbf{63.1} & \textbf{81.8} / \textbf{78.5} & 48.8 / 36.3  & \textbf{80.9} / \textbf{77.8} & \textbf{74.4} / \textbf{56.2} & 47.8 / 37.5 & 59.3 / \textbf{12.6}  & 61.3 / \textbf{12.2} & 81.1 / 23.9 \\ \bottomrule
\end{tabular}
}
\end{table}

\textbf{Scenario 1.}\ Table \ref{tab:cross-distribution-results-table} shows the error / tSuc results for all three cross-distribution scenarios. First, consider the no training data overlap case as represented by RN50-A$\rightarrow$DN121-B and RN50-B$\rightarrow$DN121-A. For baselines, we include the RN50-A$\rightarrow$DN121-A and RN50-B$\rightarrow$DN121-B. As hypothesized, the cross-distribution transfer performance is worse than the within-distribution performance. However, the \textit{FDA}-based methods still perform well, and significantly better than \textit{TMIM}. For the DN121-B target model case, \fdanlxent{5} induces $77\%$ error / $63\%$ tSuc when transferred from the RN50-A whitebox, which is an improvement over \textit{TMIM} of about $40\%$ error / $51\%$ tSuc. When compared to the \fdanl{1} method, our multi-layer extensions lead to improvements of over $9\%$ error / $17\%$ tSuc. Similar results are shown in RN50-B$\rightarrow$DN121-A.

\textbf{Scenario 2.}\ Next, we analyze the case in which the blackbox model is trained on a subset of the whitebox model's training dataset, i.e., RN50-Full$\rightarrow$DN121-A/B. Different from the previous experiment, the complexity of the whitebox and blackbox tasks is significantly different (1,000 class versus 15 class classification), where the whitebox model's task is more challenging. Remarkably, despite such a gap in the task and dataset complexity, the transfer results from the feature distribution attacks are quite high. In both cases, the error is near $60\%$ and tSuc is over $45\%$. This is as compared to the near $16\%$ error / $5\%$ tSuc from the \textit{TMIM+SGM} attack. Again, we also see large improvements using multi-intermediate-layer attacks. Interestingly, the \textit{+xent} component is not helpful in this situation. This is likely because the decision boundary structure of the two models is sufficiently different that it is no longer relevant for generating attacks.

\textbf{Scenario 3.}\ In the last case, the whitebox model is trained on a subset of the blackbox model's dataset, which is represented in the RN50-A/B/C$\rightarrow$DN121-Full columns. Not surprisingly, the transfer rates of all methods are lower in this setting, likely because the complexity of the whitebox model's task is much simpler than the blackbox model's task. Thus, the level of detail and granularity of the features learned in the whitebox model is less. However, as with the other tests, all \textit{FDA}-based methods significantly outperform the \textit{TMIM} method, and the multi-layer upgrades significantly advance the performance over the vanilla \fdanl{1}. An interesting takeaway is that the size of the whitebox's subset matters greatly. Even as we look from RN50-A/B to RN50-C as whitebox models, the performance increases by over $20\%$ error / $12\%$ tSuc.

\vspace{-2mm}
\subsection{Query-based extension} \label{sec:query-extension}
\vspace{-2mm}
A criticism of transfer-based blackbox attacks is that the success rate can be dependent on the whitebox-blackbox pair: certain combinations may have more limited transfer in some instances, as can be seen when MNv2 is the blackbox model in Table~\ref{tab:full-inet-transfer-table}. 
In contrast, query-based blackbox attacks directly estimate the gradient of the blackbox model  via repetitive querying during attack generation \cite{UesatoOKO18, IlyasEAL18, ChengDPSZ19}, and typically achieve higher success rates than transfer-based attacks. The principle drawback to query attacks, which does not afflict transfer methods, is that for each adversarial example the attacker may have to query the target model tens-of-thousands of times to properly estimate the gradient, which in some cases may not be feasible due to time and monetary constraints, or potential threat detection systems in the target model. Motivated by query efficiency, \citet{ChengDPSZ19} incorporate the adversarial direction from a transfer-based attack as a prior in the P-RGF attack. To show a potential way to extend our methods, in situations where limited queries to the blackbox model are acceptable, we investigate using \fdanlxent{5} and \textit{TMIM} methods as \textit{priors} in \citep{ChengDPSZ19}.
To integrate with P-RGF, we extend the transfer methods to perturb for 15 total iterations. In the first 10, the noise is solely optimized on the whitebox model (as usual). Only in the last 5 iterations do we use the transfer direction as the prior in the P-RGF estimator. We find warm-starting the transfer direction in this way to be effective. The tSuc results for several transfer scenarios, under different query budgets, for \textit{TMIM} / \fdanlxent{5} priors, are shown in Table \ref{tab:fda-with-prgf-table}.

\vspace{-5mm}
\begin{table}[h]
\centering
\caption{tSuc of transfer-based attacks used as a prior with P-RGF (prior = \textit{TMIM} / \fdanlxent{5})}
\label{tab:fda-with-prgf-table}
\resizebox{.8\columnwidth}{!}{
    \begin{tabular}{c|ccc|ccc}
    \toprule
               & \multicolumn{3}{c|}{Whitebox Model = RN50}  & \multicolumn{3}{c}{Whitebox Model = DN121} \\ \hline
    \# Queries \textbackslash{}\textbackslash\ Target & DN121        & VGG16bn     & MNv2        & RN50        & VGG16bn     & MNv2       \\ \hline
    0          & 3.0 / 57.9   & 1.5 / 43.5  & 0.1 / 22.9  & 2.3 / 50.1  & 1.3 / 48.0  & 1.0 / 24.4 \\
    100        & 5.5 / 70.5   & 3.3 / 60.3  & 2.1 / 35.5  & 4.8 / 63.0  & 3.1 / 65.8  & 2.0 / 37.1 \\
    500        & 7.5 / 81.2   & 6.2 / 77.3  & 4.2 / 56.8  & 6.6 / 79.7  & 5.8 / 86.2  & 3.9 / 65.8 \\
    1000       & 10.2 / 87.3  & 9.9 / 86.5  & 6.9 / 71.9  & 9.0 / 88.9  & 9.3 / 93.5  & 6.4 / 82.3 \\
    2000       & 14.9 / \textbf{92.9}  & 16.9 / \textbf{93.4} & 12.3 / \textbf{85.5} & 13.0 / \textbf{94.7} & 16.0 / \textbf{97.3} & 11.9 / \textbf{93.2} \\ \bottomrule
    \end{tabular}
}
\end{table}


\vspace{-3mm}

In all transfer cases, we observe that the \fdanlxent{5} attack yields a significantly better prior than \textit{TMIM}, which is aligned with the performance gaps observed in previous experiments. As the number of queries to the blackbox model increases, the attack success rate increases, and the margin of performance between the two priors grows to nearly $80\%$ tSuc. With only 100 queries, the attack success rate of \fdanlxent{5} increases by over $10\%$, and with 2,000 queries the average tSuc rate is over $90\%$. This result motivates future work on how to improve the feature space attack framework through the allowance of a limited budget of queries to the target model during attack generation.
\vspace{-2mm}
\section{Conclusion}
\vspace{-2mm}
We introduce a feature space-based adversarial attack framework that allows for perturbations along the extracted feature hierarchy of a DNN image classifier to achieve state-of-the-art targeted blackbox attack transferability. In the ``standard'' blackbox transfer case, where the source and target model share ImageNet-1K as a training dataset, our methods outperform output-layer-based attacks by a factor of $10\times$ and existing feature-space methods by a factor of $2-3\times$. These performance gains are attributed to the inclusion of \textit{multi-layer} information, which leads to significantly higher disruption in the feature spaces of both the whitebox and blackbox. In a set of three ``strict'' blackbox transfer scenarios, where the training dataset and label spaces of the whitebox and blackbox models differ, we show that our methods maintain similar performance margins over the baselines, even when there is no training data overlap or a 985-class discrepancy in the label space. Finally, we show an extension of the method for situations when it is acceptable to query the target model during attack generation.

\newpage
\bibliography{citations}


\newpage

\section*{Appendix}

\subsection*{A. Layer decoding tables}

\begin{figure}[h]
\centering
    \includegraphics[width=.6\linewidth,trim={0 0 0 0},clip]{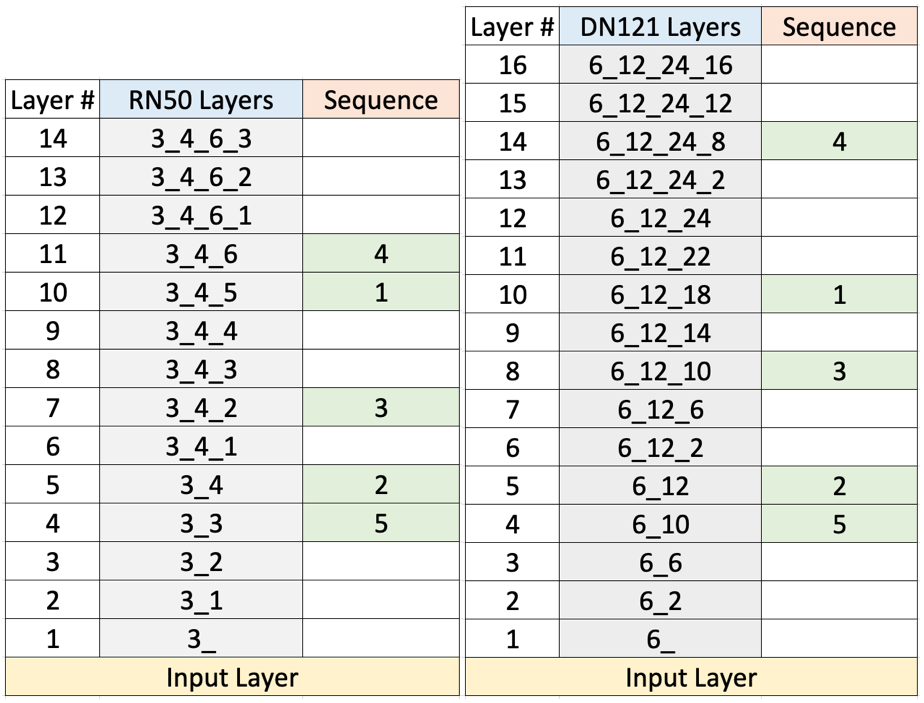}
\caption{Layer notation of whitebox models and sequence in which layers get added to multi-intermediate-layer attacks.}
\label{fig:layer-decoding}
\end{figure}

Here we discuss the DNN layer notation used throughout the work. We use two whitebox models: ResNet50 (RN50) \cite{He2016DeepRL} and DenseNet121 (DN121) \cite{Huang2017DenselyCC}, which have been shown to be good sources for generating transferable adversarial examples. 

\begin{figure}[h]
\centering
    \includegraphics[width=.99\linewidth,trim={50 0 55 0},clip]{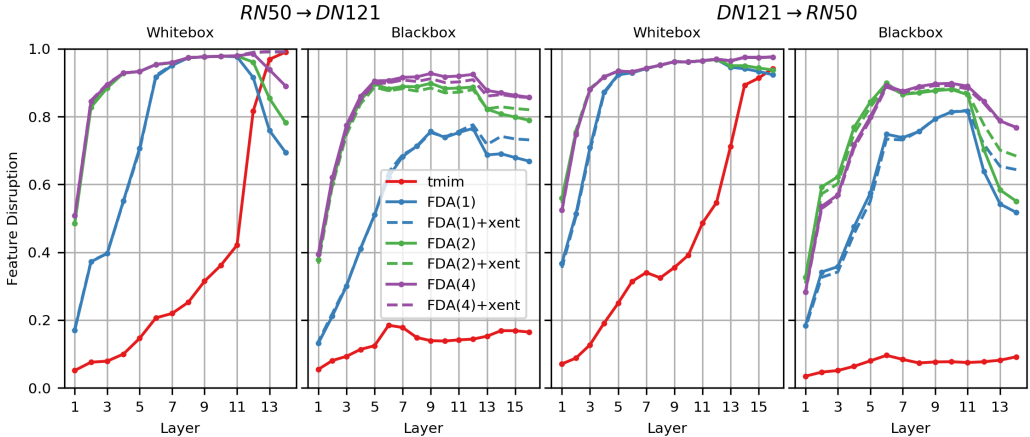}
\caption{Disruption of features caused by transfer attacks.}
\label{fig:feature-disruption-appendix}
\end{figure}

The RN50 notation follows the implementation in \url{https://github.com/pytorch/vision/blob/master/torchvision/models/resnet.py}. RN50 has 4 layer groups, with \{3, 4, 6, 3\} Bottleneck blocks in each, respectively. Of the 16 possible layers we notate 14 of them in Figure \ref{fig:layer-decoding} where ``deeper'' layers closer to the output of the model have higher layer numbers. This is the notation used in Figures \ref{fig:feature-disruption} and \ref{fig:feature-disruption-appendix}.

The DN121 notation follows the implementation in \url{https://github.com/pytorch/vision/blob/master/torchvision/models/densenet.py}. DN121 also has 4 main layer groups, with \{6, 12, 24, 16\} layers in each, respectively. From the 58 possible layers, we sample 16 layers from across the depth as shown in Figure \ref{fig:layer-decoding}.

As an example of how the layers are used in the attacks, when generating an attack from a RN50 whitebox that uses layers 5 and 10, this means that we are ``probing'' the model to extract the feature map at the output of the \{3,4\} and \{3,4,5\} layers. The ``Sequence'' column is the order in which the 5 attacking layers get added to the multi-layer attack, as found by greedy optimization. See Appendix B for more information about this process.

\newpage
\subsection*{B. Additional experimental setup details}

\textbf{Auxiliary models.}\ Using pseudo-PyTorch notation, we use the following architecture for the auxiliary models:
\texttt{[Conv(\#kernels=128, kernel\_size=3, stride=1, pad=1), ReLU, MaxPool(2,2), Conv(\#kernels=128, kernel\_size=3, stride=1, pad=1), ReLU, MaxPool(2,2), Dropout(p=0.3), linear(inputs=?, outputs=1)]}. Note that the number of nodes in the linear layer depends on the spatial size of the input feature map. We use this exact architecture regardless of layer and feature map size, for simplicity reasons.
Customizing the auxiliary architecture to each layer in the whitebox model may result in better modeling of the feature distributions at each layer, but we find this architecture to perform well enough.

Each auxiliary model is trained for 8,000 iterations with \texttt{batch\_size=64}, \texttt{momentum=0.9}, \texttt{weight\_decay=5e-4}, \texttt{initial\_learning\_rate=0.001}, and a single learning rate decay step at the 6,000$^{th}$ iteration to 0.0001. For each batch, we use a weighted sampling scheme to equalize the counts of positive and negative training samples. Note, the training of each auxiliary model can be done in a massively parallel fashion.

\textbf{Greedy layer optimization.}\ 
Selection of the whitebox model layers to target with \fdanl{N} is a combinatorial problem.
Rather than try out every possible combination, we use a greedy optimization strategy to find the best layers in Section 4, which we elaborate on here.
While this greedy approach may not find the absolute optimum set of layers, we find that it still performs quite well, with each added layer delivering significant gains in attack performance.

Note, in \cite{Inkawhich_2020_ICLR} the ``optimal'' attacking layer is found via sweeping across possible layers and using the layer that empirically has the best tSuc rate. We adopt a similar scheme here.
We only optimize to find the attack layers once. When finding the attack layers for the RN50 whitebox model we use the DN121 as a blackbox, and when finding the attack layers for DN121 we use the RN50 as a blackbox. Using a random held-out set of 200 source images from the ImageNet validation set, we generate attacks from each layer to find the best single layer to attack from. To find the best two layers to attack from, we start with the best single layer and sweep the second layer to find the best complement to the first layer. This process of using the best previous layers and finding the most complementary new layer continues until we find the best 5 layers, where we notice a performance saturation. The sequence of how the attacking layers get added is shown in the ``Sequence'' columns of Figure \ref{fig:layer-decoding}. For example, for the \fdanl{3} attack on a RN50 whitebox, the layers used in the attack are \{3,4,5\}, \{3,4\}, and \{3,4,2\}. The hyper-parameters in the objective functions are found via line search. For the RN50 whitebox, $\eta=1\mathrm{e}-5$ and $\gamma=1\mathrm{e}-4$. For the DN121 whitebox, $\eta=1\mathrm{e}-6$ and $\gamma=1\mathrm{e}-4$. For simplicity, in a multi-intermediate-layer attack all $N$ layers are equally weighted, so $\lambda_\ell = 1 / N$.

\textbf{Baseline attacks.}\ The setup and configurations of the baseline attacks largely mirror the original papers. Since the results in the \textit{TMIM+SGM} \cite{sgm_attack} paper are all tuned for un-targeted attacks, we do a similar line search on a held out dataset to find $\mathrm{decay}=0.5$ for RN50 and $\mathrm{decay}=0.4$ for DN121. For the ensemble attack \cite{LiuCLS17}, all models are weighted equally, and we include momentum in the optimization (although momentum is not discussed in the original paper) because it empirically improves the results.

All of the attacks, including new ones and the baselines use $L_\infty\ \epsilon=16/255$, $\mathrm{step\_size}=2/255$ and $\mathrm{perturb\_iters}=10$.






\newpage
\subsection*{C. Full ensemble comparison results}

\begin{figure}[h]
\centering
    \begin{subfigure}{.99\linewidth}
        \includegraphics[width=\linewidth,trim={50 40 60 0},clip]{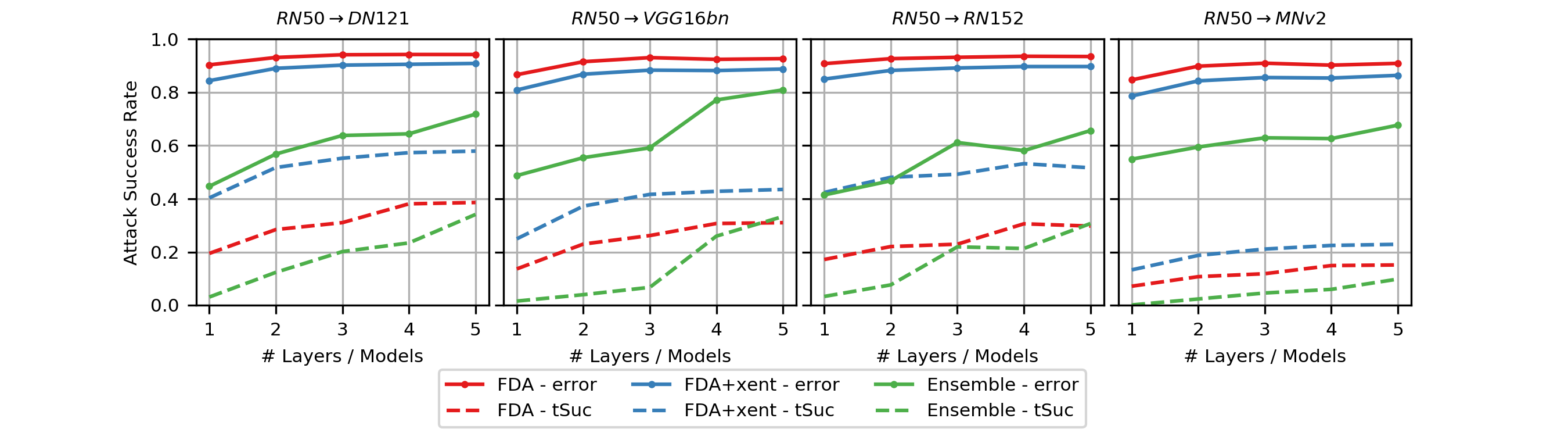}
    \end{subfigure}
    \begin{subfigure}{.99\linewidth}
        \includegraphics[width=\linewidth,trim={50 0 60 0},clip]{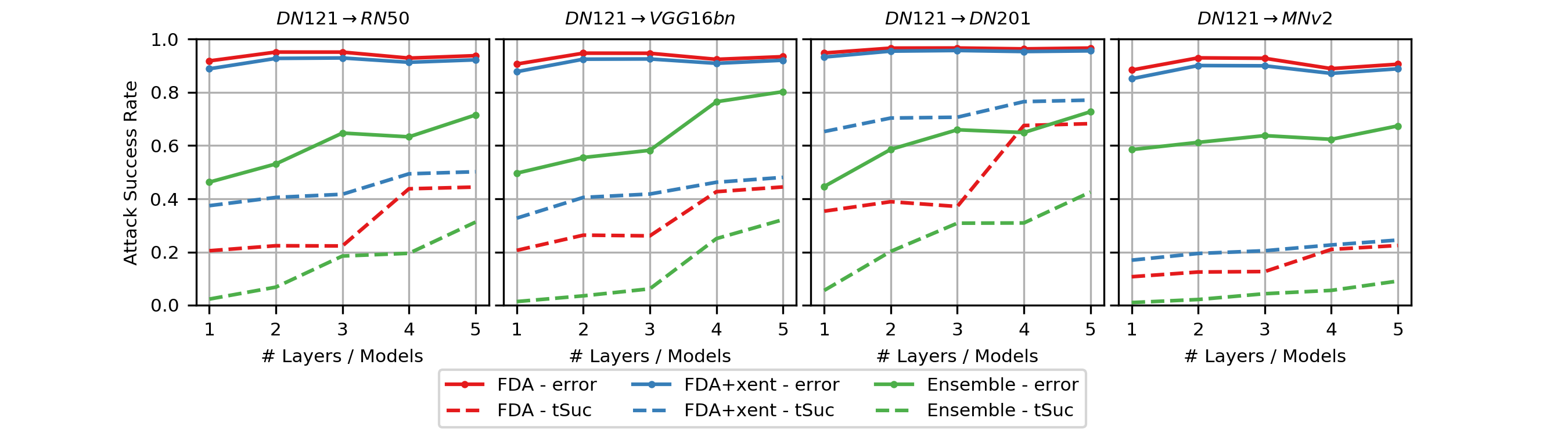}
    \end{subfigure}
\caption{Comparison of FDA methods to ensemble attacks.}
\label{fig:full-ensemble-comparison}
\end{figure}

Figure \ref{fig:full-ensemble-comparison} shown an extension of the results in Figure \ref{fig:ensemble-comparison} to include all of the transfer scenarios considered in Section \ref{sec:Inet_transfer}. The primary result is that our FDA methods, and particularly \fdanlxent{N}, outperforms the ensemble methods in all of these transfer scenarios in both error and tSuc. The critical detail is that in these plots, our method is still crafting noise using the feature space of a \textit{single} source model, where the ensemble method is using output layer information from multiple source models of varying architectures and depths. Since FDA and ensemble are seemingly orthogonal methods, we leave it to future work to explore their combination.

\newpage
\subsection*{D. Extended analysis of disruption}

Here we extend the disruption results from Section \ref{sec:disruption} to include both RN50$\rightarrow$DN121 and DN121$\rightarrow$RN50. The disruption caused by each attack is measured over 5,000 adversarial examples from each method. Adhering to the setup in Section 4, all source (clean) samples are correctly classified by both the whitebox and blackbox. The results shown here echo the conclusions drawn in Section \ref{sec:disruption}. The main conclusions are as follows:
\begin{itemize}[leftmargin=*]
    \item The \textit{TMIM} attack \cite{Dong2018BoostingAA}, which specifically leverages the output layer of the whitebox model, is only capable of causing high disruption towards the output of the whitebox model. In both the early layers of the whitebox and all throughout the blackbox model, \textit{TMIM} causes very little disruption w.r.t. the target class.
    
    \item The baseline \fdanl{1} method \cite{Inkawhich_2020_ICLR} causes significantly more disruption in the intermediate feature spaces of both the whitebox and blackbox but has two undesirable behaviors. First, the intermediate disruption is significantly higher in the whitebox than in the blackbox, even for the layers that were not considered in the attack. Ideally, the intermediate space of the blackbox model would show higher disruption. Second, the disruption in the final few layers of both the whitebox and blackbox has a sharp decrease (as best seen in the RN50 whitebox and blackbox). This indicates that although the intermediate feature maps look to be the target class with nearly $100\%$ probability, that does not automatically induce classification as the target class.
    
     \item The primary effect of the \textit{+xent} component is to create higher disruption in the final few layers, which directly addresses the weakness discussed in the previous bullet. This behavior is perhaps best illustrated by comparing the \fdanl{1} and \fdanlxent{1} results in the rightmost plot (RN50 as a blackbox). From early through middle layers, the disruption is about the same, and in the final 3 layers the \textit{+xent} version has higher disruption. As well as being quite intuitive behavior, this is evidence for why the attack performs better.
    
    \item The notable effect of using multi-intermediate-layers is significantly more disruption in the intermediate space of both whitebox and blackbox models. Looking at the whitebox models, as we go from \fdanl{1} to \fdanl{2}, the disruption is noticeably increased in the earlier layers. This is not surprising, as from Figure \ref{fig:layer-decoding}, the second attack layer to be added in both cases is a layer closer to the input layer. This trend of increased disruption in earlier layers is also observed in the blackbox models. As we move from \fdanl{2} to \fdanl{4}, from Figure \ref{fig:layer-decoding} we have added $2$ layers that are both deeper in the model than the second layer added, so we do not see increased disruption in the earlier layers. However, we do see increased disruption in the later layers, as expected because the fourth layer added is the closest to the output of the whitebox model. This pattern of how including early layers in the attack set causes early layer disruption and including late layers in the attack set causes late layer disruption matches intuition and provides insight into how the attack works.
    
\end{itemize}

\newpage
\subsection*{E. Cross-distribution experiment}

\begin{figure}[h]
\centering
    \includegraphics[width=.99\linewidth,trim={0 0 0 0},clip]{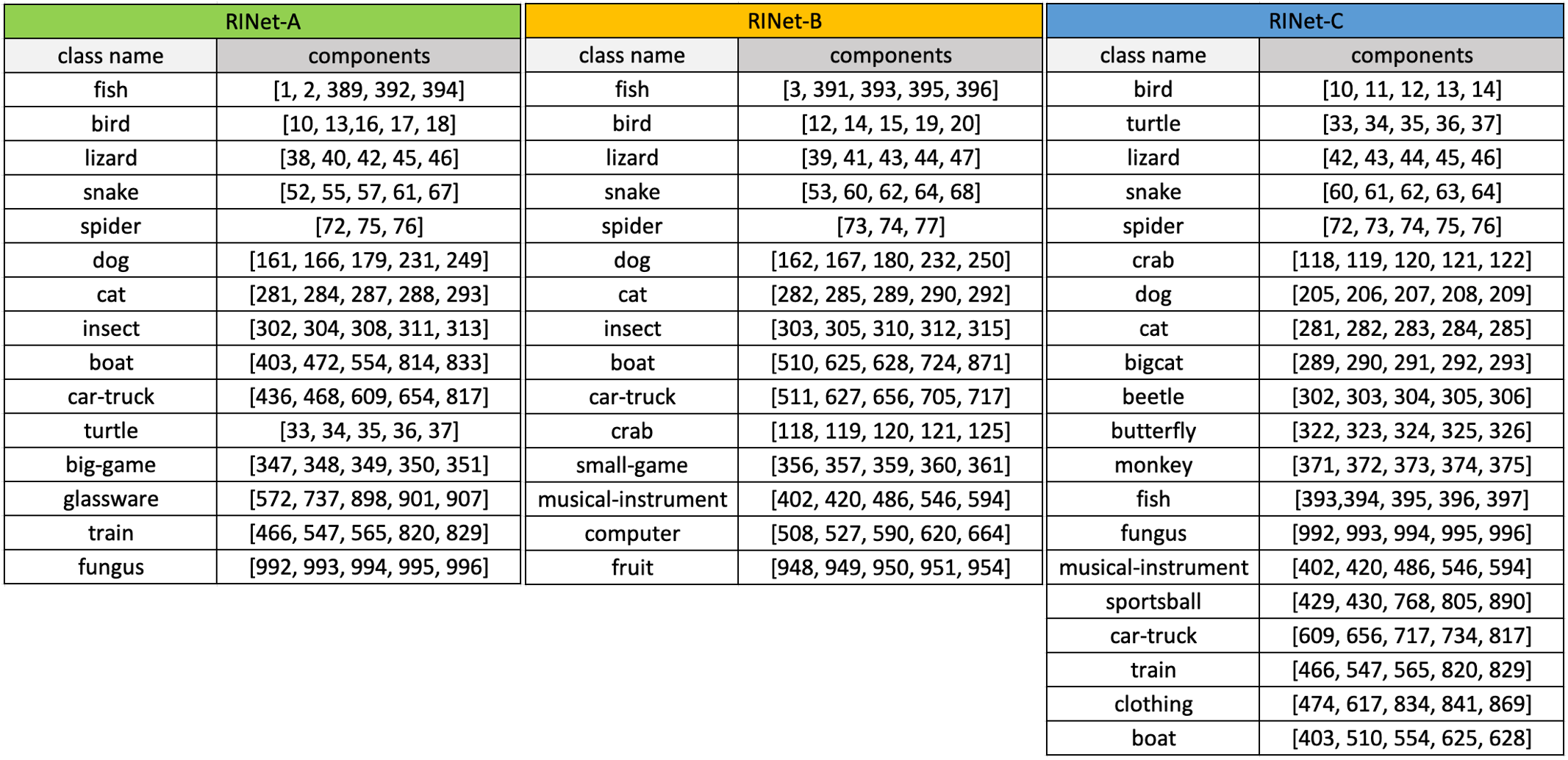}
\caption{Class splits of Restricted-ImageNet subsets.}
\label{fig:rinet-class-splits}
\end{figure}

To evaluate the effect of differences in the training data distributions of the source and target model we propose three splits of the ImageNet-1k dataset \cite{Deng2009ImageNetAL}, which we call RestrictedImageNet-A/B/C (RINet-A/B/C). To split the classes, we leverage the WordNet~\cite{miller1998wordnet} hierarchical structure of the dataset such that each class in a RINet set is a superclass category composed of multiple ImageNet-1K classes, noted in Figure \ref{fig:rinet-class-splits} as ``components.'' For example, the ``fish'' class of RINet-A (both the train and val parts) is the aggregation of ImageNet-1k classes: \texttt{[1:'tench', 2:'goldfish', 389:'barracouta', 392:'rock-beauty', 394:'sturgeon']}. See \url{https://gist.github.com/yrevar/942d3a0ac09ec9e5eb3a} for the number to category translations.

\textbf{RINet-A\&B. } One of the key experiments in this work is to measure the transfer performance when the source and target models are trained on similar categories, but have no training data overlap. This represents a more realistic attacking scenario when the adversary is aware of what classes the target model is trained on (or at least most of them) but must collect their own data as they do not have access to the target model's training dataset. To simulate this scenario, we create RINet-A and RINet-B, which each have 15 classes, of which 10 are shared between the two and 5 are unique, and have \textit{zero} training data overlap.

\begin{wrapfigure}{r}{.45\linewidth}
\vspace{-4mm}
\centering
    \includegraphics[width=\linewidth,trim={0 0 0 0},clip]{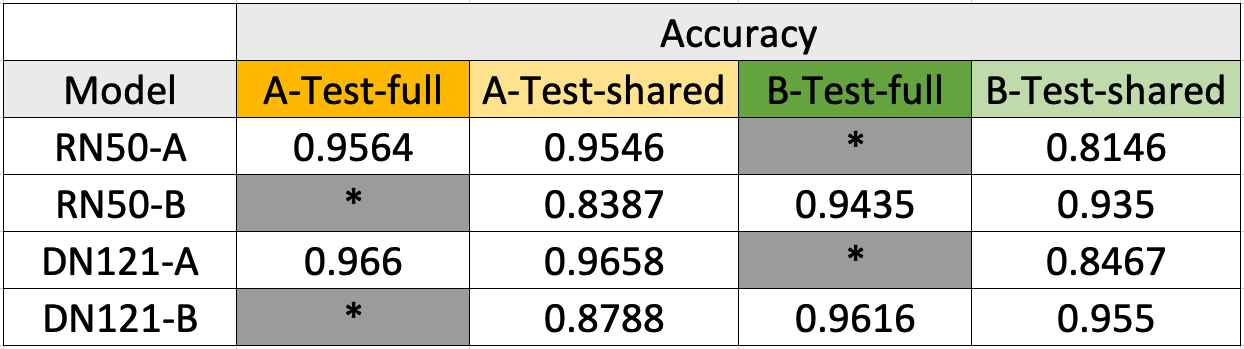}
    \caption{Standard accuracy of RINet models.}
    \label{fig:rinet-model-accuracies}
    \vspace{-4mm}
\end{wrapfigure}

To perform our experiments, we trained a RN50 and DN121 model on each of the splits using code from \url{https://github.com/pytorch/examples/blob/master/imagenet/main.py}. The accuracy of each model is shown in Figure \ref{fig:rinet-model-accuracies}. Since it is not meaningful to test an RINet-A model on the unique classes of RINet-B, we show the test accuracy on the ``full'' and ``shared'' classes only when appropriate. For the models trained and tested on the same splits, the accuracy is about $95\%$. When the models are tested on the shared class data of the other split, the test accuracy drops to about $81\%-87\%$, which is not surprising given the difference of ImageNet classes used.

\textbf{RINet-C. } The other dataset considered is RINet-C. It has very similar construction to A \& B in that each class is an aggregation of five ImageNet-1K classes. However, RINet-C has 20 classes which makes it a bigger subset than RINet-A/B. The purpose of including RINet-C is to measure how important the size of the subset is when attacking in the RN50-A/B/C$\rightarrow$DN121-FullINet and RN50-FullINet$\rightarrow$DN121-A/B/C scenarios. We train RN50 and DN121 RINet-C models in the same way as we do the RINet-A/B models. The test accuracy of RN50-C is $95.2\%$ and DN121-C is $96.3\%$.

\textbf{Cross-distribution attack settings. } Lastly, we discuss how we setup and carry out the attacks in the cross-distribution experiments. Here, we assume the auxiliary models have been trained for RN50-A/B/C/FullINet (training details in Appendix B).

\begin{itemize}[leftmargin=*]

    \item \textit{Constants across all tests: } The attacking parameters are still $L_\infty\ \epsilon=16/255$, $\mathrm{step\_size}=2/255$, and $perturb\_iters=10$.
    
    \item \textit{RN50-A/B$\rightarrow$DN121-A/B (Scenario 1): } Since the source task is significantly different than Full-ImageNet, we re-search for the best 5 attacking layers for the RN50-A/B models. We use the same greedy search technique described in Appendix B only on the RN50-A$\rightarrow$DN121-A transfer scenario. In the notation of Figure \ref{fig:layer-decoding}, the sequence of adding attacks layers in the \fdanlxent{N} framework is: \texttt{1:[3,4,6,1], 2:[3,4,6], 3:[3,4,6,2], 4:[3,4,5], 5:[3,4,4]}. We also find that increasing the weight of the cross-entropy term to $\gamma=1e-3$ helps. Other attacking parameters stay the same: $\eta=1e-5, \lambda_\ell = 1/N$. When attacking, we only consider source samples from the shared-classes that are correctly classified by both the source and target models. We then target each of the other 9 shared classes individually and do this for all source samples in the whitebox model's validation dataset. Given the classes are shared, the computation of error and tSuc is straightforward.
    
    \item \textit{RN50-A/B/C$\rightarrow$DN121-FullINet (Scenario 3): } In these tests, we use the exact layer-set and hyper-parameters used in Case 1. The only tricky part is how to define attack success because the class structures are significantly different. Here, since the whitebox model is RN50-A/B/C, we work in the label space of RINet-A/B/C. In the attack loop, for each source sample in the whitebox model's validation set that is correctly classified by both the whitebox and blackbox, we target each of the other 14 (or 19 in the case of RINet-C) classes individually. An attack is successful if the blackbox model's top-1 prediction is one of the components that makes up the RINet target class. For example, if we use RN50-A as the whitebox model, for the target label ``spider'' we count a tSuc iff the DN121-FullINet model's prediction is in $[72,75,76]$. This is also how we check that the blackbox model is initially correct for each source sample. Note, this is a somewhat restrictive definition of tSuc. If we attack with the target label ``dog,'' we only count a tSuc if the blackbox model's prediction is a component that made up the RINet's dog class. However, ImageNet-1K has over 150 ``dog'' sub-classes but tSuc only accounts for 5 of these.
    
    \item \textit{RN50-FullINet$\rightarrow$DN121-A/B/C (Scenario 2): } In these evaluations, we use the attack configurations, layer set, and hyper-parameters from the main ImageNet transfer tests (i.e., we do not tune for this particular test). As in Case 2, since the label spaces of the whitebox and blackbox models are different, we must handle the conversion as to measure attack success. For each source image in the ImageNet-1K validation set, we first check that it is a component of the target model's dataset, then we check that it is correctly classified by both the whitebox and blackbox using their respective label spaces. Using images that passed both these checks, we then target each of the other 14 (or 19) classes in the RINet label space by randomly choosing a Full-ImageNet target label from the component set of the target RINet class. For example, if we want to target ``bird'' on the RN50-B blackbox model, we would randomly select a label from the component set $[12, 14, 15, 19, 20]$ and use that as the target label in the Full-ImageNet space. With this, computing error and tSuc is straightforward.

\end{itemize}

Lastly, we would like to emphasize that we did not re-tune all of the hyper-parameters and layer sets for each individual transfer. We mention this to illustrate that the method and results are not ultra-sensitive to these parameters (within reason). However, it would not be surprising if the transfer results improve if we tuned the hyper-parameters for each situation.

\end{document}